\documentclass[a4paper,11pt]{article}

\usepackage{jinstpub} 
\usepackage{lineno}

\title{\boldmath Single event effects in the HCCStar ASICs for ITk strip upgrade}

\author[a,b,*]{Shaogang Peng,\note{Corresponding author.}}
\author[c]{Jeff Dandoy,}
\author[d]{Bruce Gallop,}
\author[e]{Hantao Jing,}
\author[f]{Jaya John John,}
\author[g]{Paul Keener,}
\author[f]{Pedro Leitao,}
\author[e]{Qiang Li,}
\author[b,h,*]{Weiguo Lu,}
\author[g]{Godwin Mayers,}
\author[g]{Mitch Newcomer,}
\author[d]{Peter Phillips,}
\author[e]{Zhixin Tan,}
\author[i]{Matt Warren,}
\author[a]{Yan Zhou}

\affiliation[a]{Department of Physics, Tsinghua University,\\Haidian District, Beijing, China}
\affiliation[b]{Institute of High Energy Physics, Chinese Academy of Sciences,\\19B Yuquan Road, Shijingshan District, Beijing, China}
\affiliation[c]{Carleton University,\\Ontario, Canada}
\affiliation[d]{Particle Physics Department, STFC Rutherford Appleton Laboratory, \\Harwell Science and Innovation Campus, Didcot, United Kingdom}
\affiliation[e]{Spallation Neutron Source Science Center, \\Dongguan, China}
\affiliation[f]{Experimental Physics Department, CERN, \\Geneva, Switzerland}
\affiliation[g]{Department of Physics and Astronomy, University of Pennsylvania, \\South 33rd Street, Philadelphia, U.S.A.}
\affiliation[h]{State Key Laboratory of Particle Detection and Electronics, \\19B Yuquan Road, Shijingshan District, Beijing, China}
\affiliation[i]{Department of Physics and Astronomy, University College London, \\Gower Street, London, United Kingdom}

\emailAdd{shaogang.peng@cern.ch, weiguo.lu@cern.ch}

\abstract{Triple Modular Redundancy (TMR) technology has been implemented in the design of the HCCStar to reduce digital state changes and ensure reliable operation.\ We tested the effectiveness of the protection by placing HCCStar chips
in a proton beam.\ We
studied corrected bit flips in registers and actual single event effects in the LCB and LP paths under
different proton energies.\ Our estimate is that the LP data-loss fraction relative to the 400 kHz readout is $\mathcal{O}(10^{-10})$ during normal operation and there will be $\mathcal{O}(10)$ corrected bit flips per HCCStar bit per year at the HL-LHC.}


\begin{document}
\maketitle
\flushbottom

\section{Introduction} 
\label{sec:intro}

To prepare for the High-Luminosity upgrade of the Large Hadron Collider (HL-LHC), the ATLAS experiment is replacing its Inner Detector with the all-silicon Inner Tracker (ITk), which includes the innermost pixel layers and the outermost strip layers~\cite{collaboration2008atlas}.\ The ITk is specifically engineered to withstand high radiation levels and operate at faster readout speeds, which are critical to manage the increased luminosity expected at the HL-LHC~\cite{dandoy2023irradiation}. 

The ABCStar, the latest version of the ATLAS Binary Chip (ABC), has undergone extensive testing to assess its radiation resistance~\cite{basso2022starry, peng2024single}.\ The Hybrid Controller Chip (HCC) plays a pivotal role by transmitting clock and control signals to the ABC and receiving data back from it.\ We tested the latest version of the HCC, HCCStar Version 1.\ In the worst-case scenario, these ASICs are anticipated to receive a maximum total radiation dose of up to 50 MRad~\cite{atlas2017technical}. 



It is essential to conduct extensive testing to ensure that the HCC
will operate reliably in the high-radiation environment of the HL-LHC.\ HCCs have undergone rigorous tests for radiation tolerance using gamma irradiation and heavy ions~\cite{dandoy2023irradiation, dandoy2023testing}.\ We have performed the irradiation of HCC with proton beams, adjusting the beam energy to explore the relationship between bit flip and energy.\ Through this beam testing, we aim to perform a comprehensive analysis of the HCC.

Single Event Effects (SEEs) are phenomena that occur when particles ionize materials as they pass through them and cause an observable change.\ Specifically, two types of SEEs to which the HCC is particularly susceptible are Single Event Upsets (SEUs) and Single Event Transients (SETs).\ SEUs occur when the charge deposited by an ionizing particle affects both dynamic and static memory registers storing logic states~\cite{karnik2004characterization}.\ On the other hand, SETs occur when a transient charge pulse generated by an ionizing event leads to temporal disorder of the signal~\cite{ferlet2013single}.\ The SEE protection mechanism built into ASIC designs must undergo thorough testing, which is achieved by exposing the ASICs to test beams in specialized laboratory environments and closely monitoring their operational performance.\ During the delivery of these beams, SEEs are anticipated to take place and are expected to be mitigated by the ASICs' protective mechanisms.\ Distinguishing between SEUs and SETs can be challenging during these beam tests; however, simulations have confirmed that the strategies adopted in the ASIC design protect against both types of events~\cite{ashmanskas2023verification}.

\section{Electronics architecture}

When charged particles traverse the silicon sensor, they deposit signal charge in the diode. That charge is routed by wire bond to the front-end ASIC (ABCStar). Each ABCStar integrates 256 channels of pre-amplification and discrimination, together with an L0 buffer, event buffer, and cluster finder logic. In operation, the strip signal on every channel is amplified and shaped, then thresholded to yield a binary hit output~\cite{atlas2017technical}.

\begin{figure}[htbp]
\centering
\includegraphics[width=0.9\textwidth]{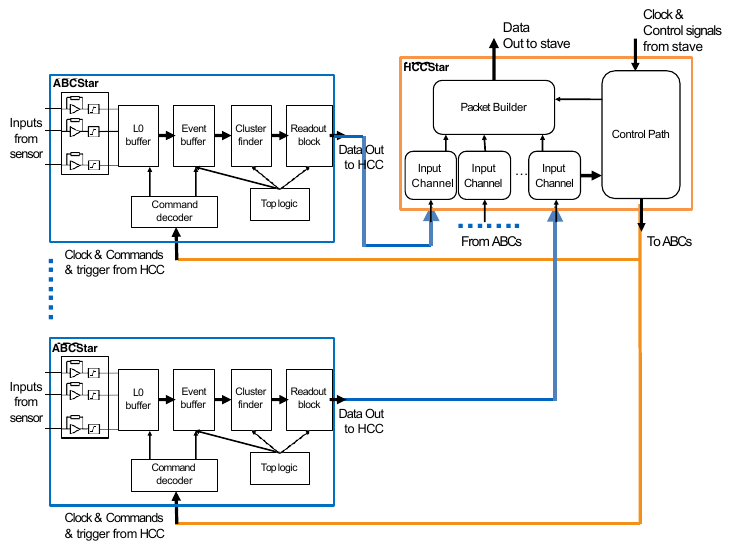}
\caption{Block diagram of a hybrid for the ITk strip modules~\cite{atlas2017technical}.\label{fig:1_1}}
\end{figure}

Up to 11 ABCStar devices are mounted on a single hybrid, depending on sensor type. One HCCStar serves each hybrid, bridging the stave or petal service infrastructure and the local ABCStar array. The HCCStar collects readout data from the ABCStars, builds event packets, and forwards them off the module; it also receives and decodes the Trigger, Timing and Control (TTC) stream and distributes clock and control to the ABCStar~\cite{atlas2017technical}. Both ABCStar and HCCStar are intended for fabrication in 130 nm CMOS. Figure ~\ref{fig:1_1} gives a schematic view of on-hybrid electronics chain.

For short strip modules, low and high voltage (LV and HV) are delivered through a dedicated power board mounted on the sensor between the two hybrids. The same board integrates a point-of-load LV converter~\cite{DCDCreview} and an HV switching circuit, and carries the Autonomous Monitor and Control (AMAC) chip, which handles module level monitoring, environmental readout, and the controlled power up sequence. 

\begin{figure}[htbp]
\centering
\includegraphics[width=0.9\textwidth]{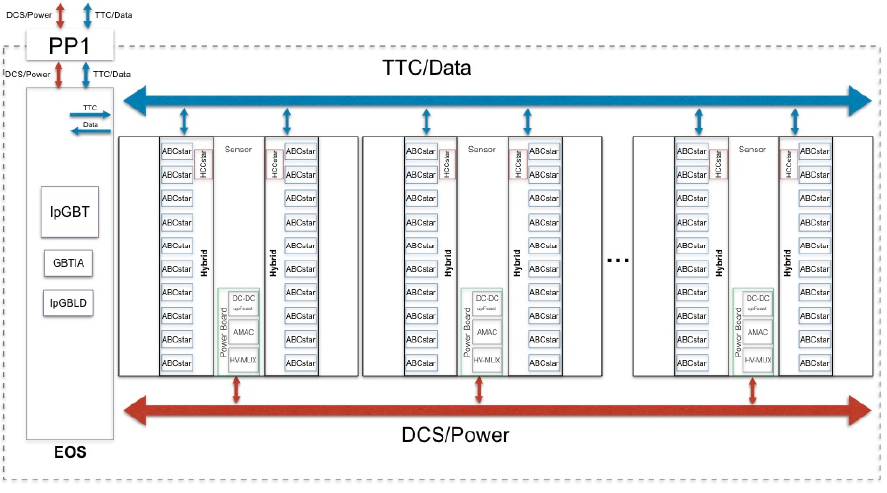}
\caption{Schematic overview of on-detector electronics for the ITk Strip Detector. Trigger, timing, and control, power distribution, and detector control system (DCS) services enter each stave or petal at the EoS card~\cite{atlas2017technical}.\label{fig:1_2}}
\end{figure}

From the End of Substructure (EoS) card, power is routed onto the stave or petal power bus and reaches each hybrid through a power interface. Figure~\ref{fig:1_2} shows the electrical layout for a short-strip stave of the ITk Strip Detector power and readout system.

Off-detector TTC signals are brought onto the detector at the EoS card and distributed to each HCCStar via the TTC bus on the bus-tape. The TTC signal stream consists of a 40 MHz system clock, serial commands and L0 triggers, and a separate R3/L1 trigger. TTC bus, readout data, and power bus share one copper-kapton bus tape, co-cured with the stave core. 

At the EoS, a low power GigaBit Transceiver (lpGBTx) provides the electrical interface to the HCC, together with a Versatile Link receiver (VTRx+) for the fibre optic driver~\cite{versatilelink}.

\section{HCC chip}

The HCC ASIC serves as the digital interface for the ITk Strips hybrids in both the endcap and barrel.\ It receives a 160 MHz clock signal and two 160 Mbps control signals over a multidrop line originating from the End of Substructure card (EoS).\ The HCC decodes the control data and generates a 40 MHz bunch crossing clock by dividing the 160 MHz clock and setting its phase from the control data framing.\ The HCC distributes this clock along with relevant control signals to between 6 and 11 ABCStar ASICs on the hybrid via shared multidrop lines.\ These control signals include triggers, event data readout requests, register read and write commands, test pulse generation, and reset signals.

\begin{figure}[htbp]
\centering
\includegraphics[width=1.0\textwidth]{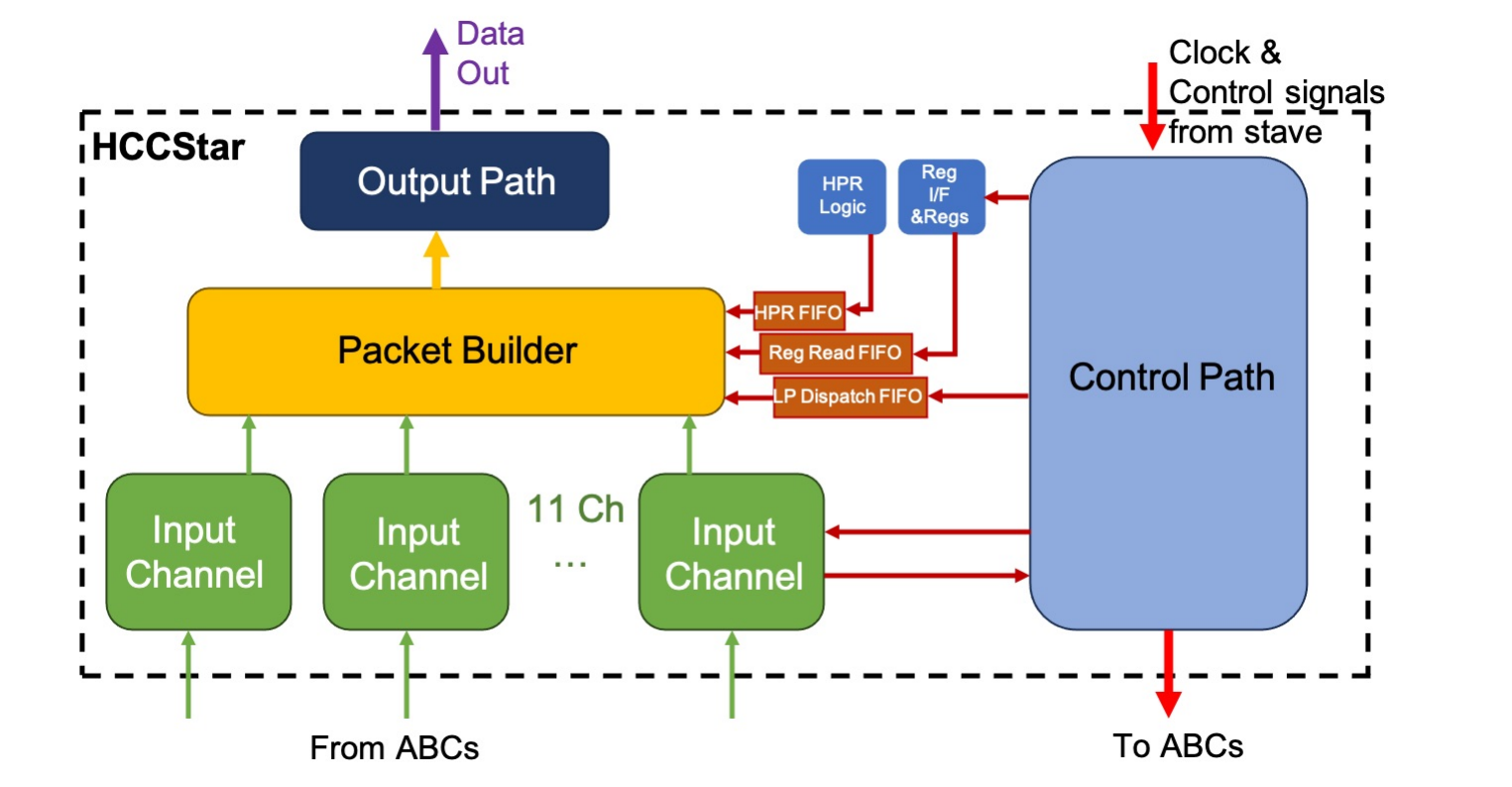}
\caption{HCC Top Level Block Diagram~\cite{atlas2017technical}.\label{fig:1}}
\end{figure}

The top level block diagram of the HCC is illustrated in Figure ~\ref{fig:1}.\ System clock and control signals enter the HCC from the stave and pass through the Control Path, where these signals are processed and modified as necessary for the ABCStar ASICs on the hybrid.\ Data from the ABCStar ASICs on the hybrid is received by the HCC’s Input Channels, with one dedicated channel for each ABCStar.\ The Packet Builder then collects data from the various Input Channels, combines them into event fragments, serializes the combined data, and sends the resulting packets across the bus tape to the EoS.

\section{CSNS irradiation}

In September 2023, tests were performed at the Associated Proton Experimental Platform at CSNS~\cite{zhang2024radiation} in China, where two HCCs were simultaneously exposed to 20 \textasciitilde{} 80 MeV protons.\ 
The flux were from $4.64 \times 10^{7}$  to $5.32 \times 10^{9}$ p/cm$^{2}$/s, the total fluence were about $7.06 \times 10^{14}$ p/cm$^{2}$, and the total dose was about 24 MRad.\ To minimize the impact of SEE on the circuits as much as possible, two HCC setups were placed in the lead bricks bunker, excluding the single-chip board.\ Over 37~h of beam time, the ASICs experienced a total fluence equivalent to approximately 48\% of the HL-LHC's expected lifetime.

The HCC chips utilize the Inner Tracker Strips DAQ (ITSDAQ) software for data readout.\ The ITSDAQ firmware is loaded on the Nexys Video FPGA board, enabling monitoring of both register and physics data.\ However, the buffers that receive data from the ABCs do not protect the physics data path from SEEs, except for metadata like L0tag, BCID, and the last cluster bit which are secured.

When Triple Modular Redundancy (TMR) is enabled, any changes in the data bits are detected by the triple modular circuit.\ This setup compares the altered bits with two other circuits to decide whether to retain the changes.\ Following this comparison, the bits are reset, ensuring they do not disrupt the ASICs’ normal functions.\ The corrected bits in this context are referred to as Corrected Bit Flips (CBFs).

Conversely, in circuits lacking TMR protection, actual SEEs are observed directly, without any CBFs.\ These SEEs can happen in both the registers and the physics data path, impacting components like the L0A/CMD/BCR (LCB) or Low Priority readout request (LP) lines.\ To specifically address SEEs that occur in registers (ASIR), dedicated registers (address 0, 1) count the occurrences in other registers.\ The LCB line transmits messages to the HCC, while the LP line relays the trigger decision from the HCC to the ABCs.\ Although SEEs in the LP path (ASLP) were detected, no SEEs were observed in the LCB data line during our testing.

\section{Results}
For the registers, we need to normalize according to the number of bits monitored for each individual chip.\ The calculation formula for the cross section of SEEs in the registers is shown as: 

\begin{equation}
\label{eq:x}
\begin{aligned}
\sigma_{\mathrm{reg}} &= \frac{\#\text{Bit Flips (CBFs or actual SEEs)}}{\#\text{fluence} \cdot \#\text{bits monitored}} \,,
\qquad
\end{aligned}
\end{equation}
we monitored a total of 16 registers $\times$ 32 bits/register = 512 bits.\ When calculating the cross section for physics data, the formula is as follows:

\begin{equation}
\label{eq:x1}
\begin{aligned}
\sigma_{\mathrm{LCB\ or\ LP}} &= \frac{\#\text{Bit Flips (actual SEEs)}}{\#\text{fluence}} \,,
\qquad
\end{aligned}
\end{equation}
in the case of zero measured bit flip, 95\% upper confidence bound is calculated by assuming 3 bit flips, we assume that the observed bit flips follow Poisson statistics distribution.

The process of CBFs is continuous, leading to a difference in the fluence of CBFs observed in registers compared to those detected in the data path, since we are able to monitor bit flips that are being continuously corrected when accessing the SEE registers.\ In contrast, sensitivity to SEEs within the data path is limited to the short period during data readout.

To test the effectiveness of TMR, we placed two chips simultaneously in a 70 MeV beam and performed irradiation tests with and without triplication enabled, the irradiation time was kept the same for both chips.\ The cumulative cross section for each chip is shown in Table~\ref{tab:t1}.
\begin{table}[htbp]
\centering
\caption{Summary of cumulative SEE cross sections from CSNS proton irradiation at 70~MeV. Corrected bit flips (CBFs) are observed in triplicated registers with TMR enabled; Actual SEEs in register (ASIR) and LP control-path events (ASLP) are reported with TMR disabled. Irradiations were performed at the CSNS Associated Proton Experimental Platform with a beam spot diameter of approximately 20~mm in 2023. Chips~154 and~152 were exposed simultaneously in the same beam sessions. The two chips were configured with TMR enabled (Triplication On) or disabled (Off) as indicated. The effective monitoring intervals were $\sim$2.1~h. Where no events were observed, the cross sections are 95\% confidence-level upper limits calculated assuming three events (Poisson statistics).\label{tab:t1}}
\smallskip
\small{
\setlength{\tabcolsep}{2mm}
\begin{tabular}{l|c|c|c|c|c|c|c}
\hline

Chip & Triplication & CBFs &  $\sigma_{\mathrm{CBFs}}$ [cm$^{2}$/p/bit] & ASIR & $\sigma_{\mathrm{ASIR}}$ [cm$^{2}$/p/bit] & ASLP & $\sigma_{\mathrm{ASLP}}$ [cm$^{2}$/p] \\
\hline
154  & On   & 445 & 1.02$\times$10$^{-13}$  &    0 & 6.89$\times$10$^{-16}$ &  0 & 3.53$\times$10$^{-13}$ \\
154  & Off  &  0  & 9.99$\times$10$^{-16}$  & 7953 & 2.65$\times$10$^{-12}$ & 32 & 5.46$\times$10$^{-12}$ \\
152  & On   & 450 & 1.03$\times$10$^{-13}$  &    0 & 6.89$\times$10$^{-16}$ & 13 & 1.53$\times$10$^{-12}$ \\
152  & Off  &  0  & 6.52$\times$10$^{-16}$  &  168 & 3.65$\times$10$^{-14}$ & 15 & 1.67$\times$10$^{-12}$ \\
\hline
\end{tabular}}
\end{table}

From Table~\ref{tab:t1}, we can see that there are no CBFs without triplication, moreover, the cross section when the triplication is not enabled is several orders of magnitude larger than when it is enabled, which means TMR works as expected.\ Additionally, the cross section of CBFs is approximately 10$^{-13}$ cm$^{2}$/p/bit, which is consistent with the results from TRIUMF~\cite{dandoy2023irradiation}.

With triplication enabled, ASLP was reduced to zero for chip 154 but 13 events were still observed for chip 152, comparable to the value without triplication on that chip (15 events), indicating chip-to-chip variation in the ASLP response to TMR.

\begin{figure}[htbp]
\centering
\begin{minipage}{.51\textwidth}
  \centering
  \includegraphics[width=\linewidth]{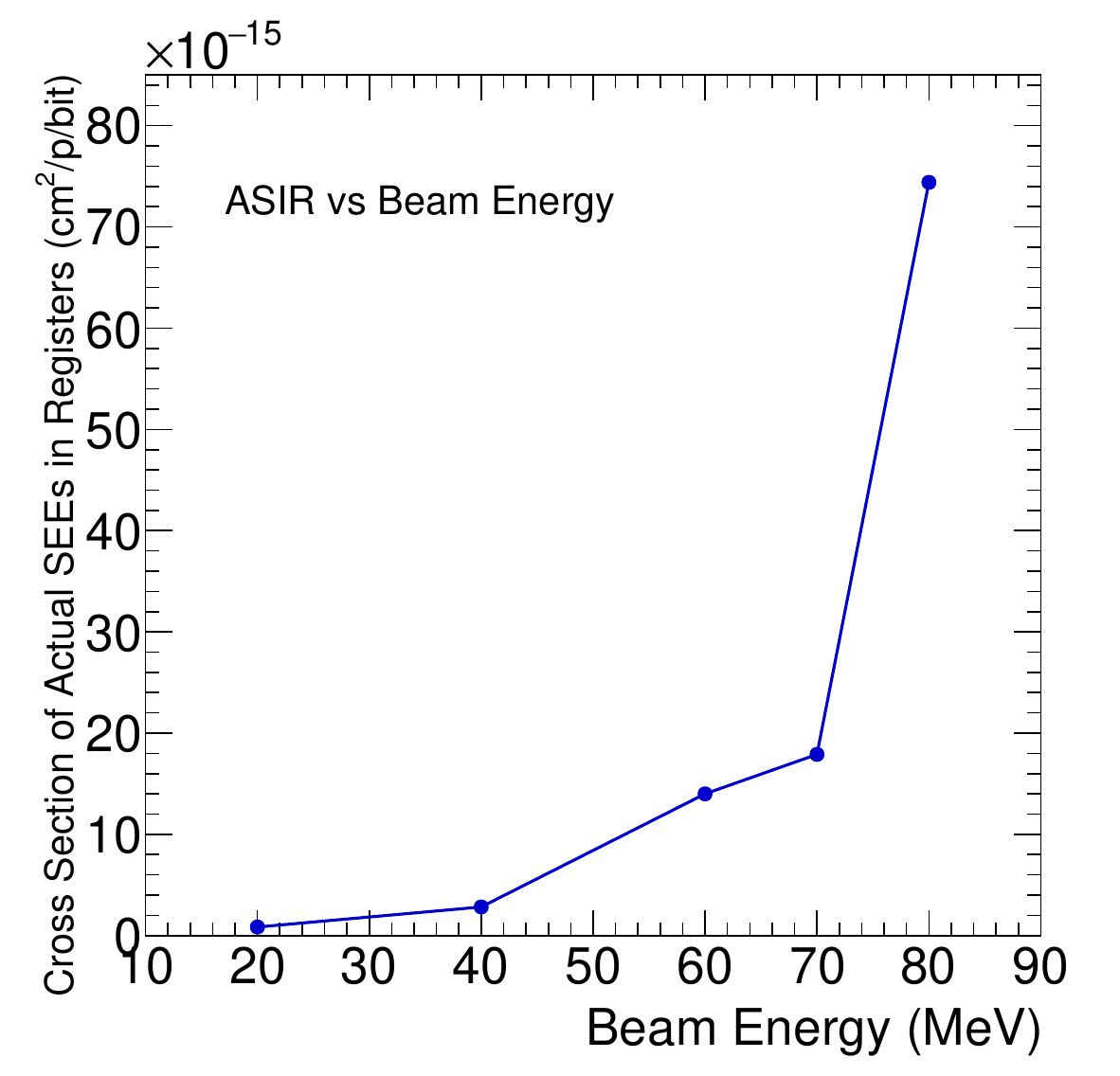}
  \text{(a)}
\end{minipage}%
\begin{minipage}{.51\textwidth}
  \centering
  \includegraphics[width=\linewidth]{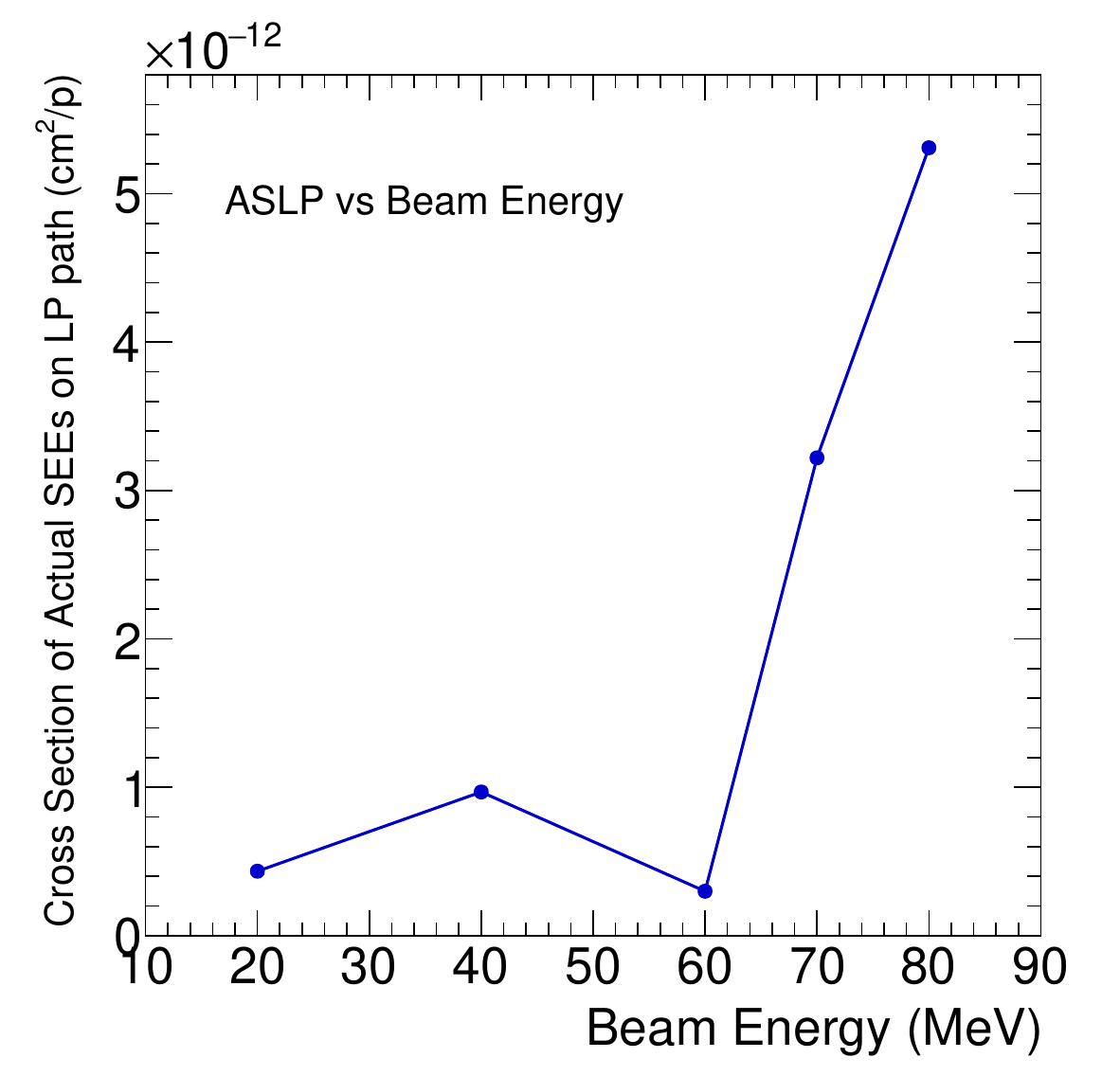}
  \text{(b)}
\end{minipage}
\caption{The cross section of actual SEEs in registers (a) or LP path (b) measured at the CSNS proton testbeam as a function of beam energy.\label{fig:2}}

\end{figure}

We also studied the cross section with beam energy varying from 20 MeV to 80 MeV.\ As seen in Figure ~\ref{fig:2}, the number of actual SEEs generally tends to increase as the energy increases. The dramatic rise in cross section from 60 MeV onward is consistent with stronger effective energy deposition, possibly reflecting a greater contribution from nuclear interactions at higher proton energies. The energy dependence is qualitatively analogous to the LET-dependent Weibull response~\cite{ecss2010hb1012}. 

We can calculate the rate of CBFs, Rate$_{\text{CBFs}}$ is given by:

\begin{equation}
\label{eq:x2}
\begin{aligned}
\text{Rate}_{\text{CBFs}} 
  &= \Phi_{\text{hadrons}} \cdot \frac{\sigma_{\mathrm{SEE}}}{\text{chip}} \cdot \frac{10^{7} \text{s} }{\text{year}} \\
  &= \mathcal{O}(10^{7}) \frac{\text{hadrons}}{\text{cm}^{2} \cdot \text{s}} \cdot \frac{\sigma_{\mathrm{SEE}}}{\text{chip}} \cdot \frac{10^{7} \text{s} }{\text{year}} \\
  &= \mathcal{O}(10^{7}) \frac{\text{hadrons}}{\text{cm}^{2} \cdot \text{s}} \cdot \mathcal{O}(10^{-13}) \frac{\text{cm}^{2}}{\text{bit}} \cdot \frac{10^{7} \text{s}}{\text{year}} \\
  &= \mathcal{O}(10) \text{CBFs/year/bit}.
\end{aligned}
\end{equation}
The hadron flux $\Phi_{\text{hadrons}}$ is $\mathcal{O}(10^{7})$ hadrons/cm$^{2}$/s, and we assume that ATLAS runs continuously for 4 months each year~\cite{basso2022starry}.

In normal operation, the HCC issues low priority (LP) readout requests on the control path, each request carries a 7-bit L0tag that identifies the event to be read from the ABCStar event buffer. For every request, the HCC tracks the corresponding ``true'' L0tag and BCID and checks the ABCStar response headers. ABCStar packets (68 bits) with an L0tag mismatch are dropped, whereas packets with a set ABCStar error flag or with a BCID mismatch are processed as usual, input-channel timeouts are likewise reported in the error block and are not described as causing packet discard. We therefore associate uncorrected SEEs on the LP control path (ASLP events, Table~\ref{tab:t1}) with at most one affected readout transaction per event and use this as a conservative upper bound on the loss or corruption of ABCStar cluster data for that hybrid channel. 

For the data-loss estimate we use the cross section from chip 154 without triplication ($\sigma_{\mathrm{ASLP}} = 5.46\times10^{-12} \mathrm{cm^2/p}$) as a conservative upper bound, which is the worst-case measured cross section in Table~\ref{tab:t1}. The corresponding upper bound on the data-loss rate per HCC is
\begin{equation}
  \mathrm{Rate}_{\mathrm{data\ loss}}^{\mathrm{(UL)}}
  = \Phi_{\mathrm{hadrons}}\,\sigma_{\mathrm{ASLP}}
  \sim \mathcal{O}(10^{-5})\,\mathrm{s}^{-1}\ \mathrm{per\ HCC},
\end{equation}
which corresponds to 546 LP-related readout anomalies per HCC per year. In the baseline ITk strip trigger scheme, the design LP readout rate is 400 kHz. For $\sim 10^7$ s of annual operation this is $\sim 4\times 10^{12}$ LP readout transactions per HCC per year, giving a fraction $\sim 10^{-10}$ relative to normal LP operation. With triplication enabled in operation, the actual rate is expected to be smaller than this conservative upper bound. Physics data loss from LP path SEEs is therefore negligible compared with normal readout operation.

\section{Conclusion}

The irradiation test results demonstrate that the triplication strategy implemented in the ASIC designs for the HCC effectively reduces SEEs. Using a conservative upper limit ASLP cross section for the LP path, it is estimated that there are 546 LP-related readout anomalies per HCC per year, implying that the LP data-loss fraction relative to the 400~kHz readout is $\mathcal{O}(10^{-10})$ during normal operation. No chip failure was observed during irradiation. Based on these results, it is anticipated that the HCC will perform well in the HL-LHC environment, with approximately $\mathcal{O}(10)$ corrected bit flips per bit per year. Bit flips in the HCC’s non-triplicated data paths are expected to be infrequent to the point of being negligible.\

\acknowledgments

This research has received funding support from the Tsinghua University Initiative Scientific Research Program and the National Natural Science Foundation of China (No.\ 11961141014).\ The research was supported and financed in large part by the National Key Research and Development Program of China under Grant No. 2023YFA1605902 from the Ministry of Science and Technology.




\end{document}